\documentclass[12pt, a4paper, parskip]{scrartcl}
\usepackage[hidelinks]{hyperref}
\usepackage[english]{babel}
\usepackage[utf8]{inputenc}
\usepackage[T1]{fontenc}  
\usepackage{lmodern, textcomp}
\usepackage[round]{natbib}
\usepackage{authblk, doi, graphicx}
\usepackage[separate-uncertainty=true]{siunitx}
\usepackage{amsmath, xspace}
\usepackage[format=plain, indention=0.5cm, font=small]{caption}
\usepackage{xcolor}

\catcode`\^^M=10
\newcolumntype{L}[1]{>{\raggedright\arraybackslash}p{#1}}
\newcommand{\ind}[1]{_\text{#1}}
\newcommand{\parent}[1]{\ensuremath{\left(#1\right)}\xspace}
\newcommand{\fof}{\!\parent}
\newcommand{\vp}{\ensuremath{v\ind P}\xspace}
\newcommand{\vs}{\ensuremath{v\ind S}\xspace}
\newcommand{\vpvsratio}{$\vp{/}\vs$ ratio\xspace}
\newcommand{\tmax}{\ensuremath{t\ind{max}}\xspace}
\newcommand{\tmaxi}[1]{\ensuremath{t_{\text{max},#1}}\xspace}
\newcommand{\ttr}{\ensuremath{t\ind{travel}}\xspace}
\newcommand{\dto}[1]{\ensuremath{\Delta t_{\text{o},#1}}\xspace}
\DeclareMathOperator{\sgn}{sgn}

\title{\vspace{-1.5cm}Toward source region tomography with inter-source interferometry: Shear wave velocity from 2018 West Bohemia swarm earthquakes}
\author{Tom Eulenfeld}
\affil{\small Friedrich Schiller University Jena, Institute for Geosciences\\Burgweg 12, 07749 Jena, Germany\\tom.eulenfeld@uni-jena.de}
\date{{\vspace{-0.8cm}\small September 08, 2020}}
\hypersetup{pdfauthor={Tom Eulenfeld}, pdftitle={Towards source region tomography}}
\defcitealias{WebnetData}{Institute of Geophysics, ASCR, 1991}

\begin{document}
\maketitle
\vspace{-1.2cm}
{\color{gray} \footnotesize
\begin{tabular}[b]{l}
An edited version of this paper was published by AGU. Copyright 2020 American Geophysical Union.\\
T. Eulenfeld (2020), \emph{Journal of Geophysical Research: Solid Earth, 125}, e2020JB019931,\\
\hspace{9.3em}doi: \href{http://dx.doi.org/10.1029/2020JB019931}{10.1029/2020JB019931}.
\end{tabular}
}
\vspace{-0.5cm}
\begin{abstract}\noindent
The concept of seismic interferometry embraces the construction of waves traveling between receivers or sources with cross-correlation techniques. In the present study cross-correlations of coda waves are used to measure travel times of shear waves between earthquake locations for five event clusters of the 2018 West Bohemia earthquake swarm.
With the help of a high quality earthquake catalog, I was able to determine the shear wave velocity in the region of the five clusters separately. The shear wave velocities range between \SI{3.5}{km/s} and \SI{4.2}{km/s}. The resolution of this novel method is given by the extent of the clusters and better than for a comparable classical tomography. It is suggested to use the method in a tomographic inversion and map the shear wave velocity in the source region with unprecedented resolution.
Furthermore, the influence of focal mechanisms and the attenuation properties on the polarity and location of the maxima in the cross-correlation functions is discussed. The intra-cluster ratio of P wave to S wave velocity is approximately fixed at 1.68.
\newline
\newline
Key points:
\begin{itemize}
\item Shear wave travel times between earthquake locations from cross-correlations of coda waves
\item Spatial variability of seismic velocity at region of 2018 earthquake swarm
\item Enhanced spatial resolution compared to classical tomography
\end{itemize}
\vspace{1ex}
\end{abstract}

\section{Introduction}
\label{sec:intro}

The  concept of seismic interferometry is most often applied by cross-correlating a coherent or incoherent seismic signal at different stations. In the limit of a sufficiently random wave field, the cross correlation function converges towards the Green's function between the two receivers \citep{Weaver2002, Snieder2004, Shapiro2004}. 
In this inter-receiver setting, the Green's function or the cross-correlation function for arbitrary time-invariant noise sources can be used to monitor the wave velocity between and near the receivers \citep[e.g.][]{Sens-Schonfelder2006, Wegler2007, Brenguier2008, Richter2014, Hillers2015, Hobiger2016, Sens-Schonfelder2019}. 
In the reverse, seismic interferometry can be applied to use earthquakes as virtual receivers at depth. In contrast to inter-receiver interferometry this approach cannot be easily used for monitoring, but has the advantage that the medium between and near the event locations is probed.\par

A number of studies used coda cross-correlations of closely spaced earthquakes to determine the inter-event travel time from the decorrelation of the coda wave field of two earthquakes \citep[referred to as method~A,][]{Snieder2005, Robinson2011, Robinson2013, Zhao2016, Zhao2019}. For this method, the earthquakes need to be separated by less than the dominant wavelength. Additionally, both events need to have the same moment tensor and the scatterer density is assumed to be constant in all directions. \cite{Snieder2005} show that under these conditions the decorrelation of the coda wave field is proportional to the earthquake separation, because both are proportional to the variance of travel time perturbations associated with different wave paths. The proportionality factor is different for different types of sources. 

\par

\cite{Hong2006} used spatial reciprocity \cite[p.~28]{Aki2002} to construct virtual pseudo-noise waveforms at each earthquake location by summing coda waves registered at different stations. These virtual recordings at different earthquake positions were then cross-correlated to obtain inter-event shear wave travel times.
\cite{Curtis2009} further demonstrated the principle of inter-source interferometry by comparing real and virtual recordings of the Sichuan earthquake.
Different from \cite{Hong2006} they used the full seismic recording to mainly reconstruct surface waves and therefore can restrict the used stations to the stationary phase zone.
\cite{Curtis2009} stated that this approach was superior in their application because more energy was used by the cross-correlation and the virtual recordings therefore were better defined.
\cite{Tonegawa2010} selected recordings in the stationary phase zone of body waves propagation between deep earthquakes and extended this inter-source interferometry concept to P and S body waves.

Inter-source interferometry as in \cite{Curtis2009} and \cite{Tonegawa2010} using direct body waves or surface waves recorded by stations located in the stationary phase zone of the two earthquakes is in the following referred to as method~B.

\begin{figure}
\centering
\newlength\q
\setlength\q{\dimexpr .32\textwidth -2\tabcolsep}
\newlength\qq
\setlength\qq{\dimexpr 0.12\textwidth -2\tabcolsep}
\renewcommand{\arraystretch}{1.5}
\hspace*{-0.08\textwidth}\begin{tabular}{L{\qq}L{\q}L{\q}L{\q}}
&Method A&Method B&Method C\\
\end{tabular}
\hfill\includegraphics[width=0.95\textwidth]{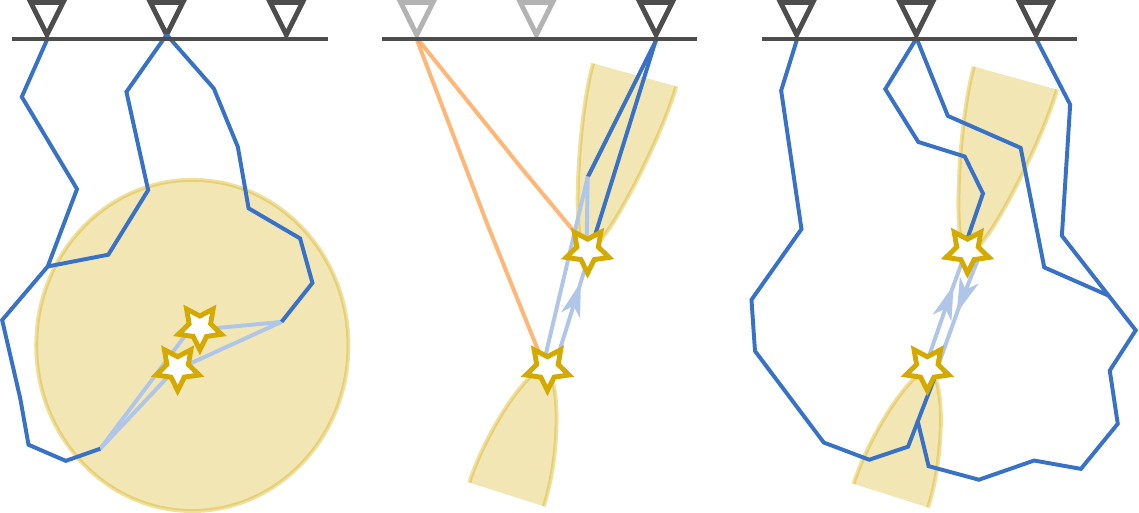}
\scriptsize
\hspace*{-0.08\textwidth}\begin{tabular}{L{\qq}p{\q}p{\q}p{\q}}
Method & Decorrelation of coda waves of nearby events & Cross-correlation of surface waves, direct P or S wavelet in stationary phase zone & Cross-correlation of inter-event coda waves\\
Event separation & Smaller than dominant wavelength& Larger than dominant wavelength & Larger than dominant wavelength\\
Station constraints & None, single channel can be used & Need to be located in stationary phase zone & None, stacking over several stations often necessary\\
Studies & \cite{Snieder2005, Robinson2011, Robinson2013, Zhao2016, Zhao2019} & \cite{Curtis2009, Tonegawa2010}, this study (using all stations) & \cite{Sun2016}, this study\\
\end{tabular}

\caption{Comparison of different methods for inter-event seismic interferometry. At the top, several ray paths of scattered and direct waves are displayed for the three methods (A, B, C). The zones of stationary phase are indicated with yellow areas. The ray paths in blue give rise to the peak in the cross-correlation function -- dark blue parts are the same for both events, light blue parts are different for both events and determine the travel time difference. In method B, light orange paths give rise to the peak in the cross-correlation function but are usually not considered because the corresponding station is not located in the stationary phase zone. On the bottom, important aspects of each method are listed.}
\label{fig:methods}
\end{figure}

\cite{Sun2016} used coda cross-correlograms to determine the S wave travel time between aftershocks of the Lushan Earthquake for an event relocation procedure (referred to as method~C). \cite{Zang2014} used autocorrelograms of coda waves to determine the event depth from the travel time of waves passing though the event location after a reflection at the surface.
Figure~\ref{fig:methods} compares the three introduced methods of inter-event seismic interferometry.

\par

\begin{figure}
\centering
\includegraphics[width=0.8\textwidth]{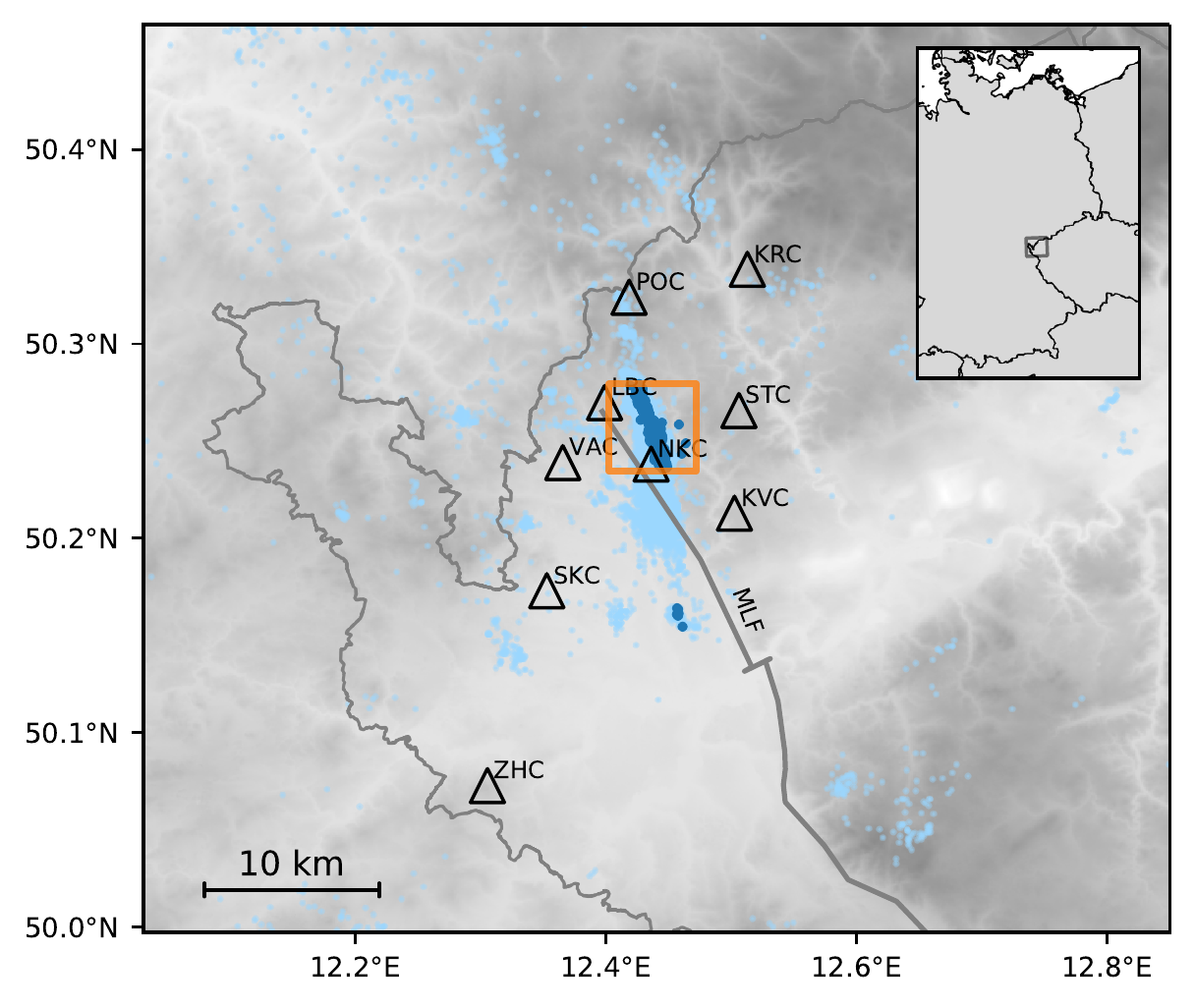}
\caption{Topographic map of Czech-German border region near Nový Kostel. Displayed are WEBNET seismic stations used in this study (triangles) together with local seismicity between the years 2000 and 2020 according to WEBNET catalog (light blue) and epicenters of 2018 Nový Kostel swarm earthquakes (dark blue). The orange rectangle defines the area of the map in figure~\ref{fig:eventmap}. The Mariánské Lázně fault zone (MLF) is indicated with a gray line.
}
\label{fig:map}
\end{figure}

In this article I further explore method~C and apply the concept of inter-source interferometry by correlation of scattered coda waves of different earthquakes. It is assumed that event locations are separated by more than a wavelength. Therefore only waves scattered in the stationary phase zone behind the two source locations interfere constructively in the cross-correlation function. The wave path are indicated in figure~\ref{fig:methods} in the right panel: Waves radiate from one event through the other event location. Waves from both events are scattered for the first time in the stationary phase zone and are recorded in the coda window with a time difference corresponding to the inter-event travel time. Waves from both sources scattered for the first time in a region outside of the stationary phase zone interfere destructively in the cross-correlation function. Therefore the maxima in the cross-correlation correspond to the inter-event travel time. This concept is basically the reciprocity of inter-station interferometry using earthquake coda waves \citep{Snieder2004} and equally applicable for isotropic scattering and the non-isotropic (forward) scattering expected in this study. 
Because more energy is radiated as S waves than as P waves \citep[approximately factor~20, e.g.][p.~188, eq.~6.10]{Sato2012} the coda will be dominated by waves originating initially from S waves. Therefore the focus is put on the determination of inter-event travel times of \emph{shear waves}.

Method~C is applied to the 2018 earthquake swarm near Nový Kostel in the Czech-German border region West Bohemia/Vogtland. The area  is of great interest as it hosts several signs of geodynamic activity such as CO$_2$ degassing, Quaternary volcanoes and earthquake swarms \citep{Fischer2014}.
The topography of West Bohemia and Vogtland is shown in figure~\ref{fig:map} together with WEBNET seismic stations and earthquake epicenters between the years 2000 and 2020.

Travel times of seismic waves between earthquake locations are of great interest as their knowledge allows to estimate the seismic velocity in the source region with a resolution determined by the earthquake distribution. Given the high event density in the volume of the swarm earthquakes this allows for a spatial resolution that is superior to conventional seismic tomography.
\cite{Lin2007} developed a double difference arrival method to determine the ratio of P wave velocity to S wave velocity (\vpvsratio) for a region embraced by a cluster of earthquake.

\cite{Dahm2013} applied a similar method to the West Bohemia earthquake swarms of the years 1997, 2000 and 2008 and found a time-dependent \vpvsratio between 1.38 and 1.70.
\cite{Bachura2016a} showed that the \vpvsratio inside different clusters of the 2014 earthquake swarm ranged from 1.59 to 1.73.

In this study the S wave velocity in the source region is determined directly from the travel time observations using a high quality earthquake catalog.
This approach is different from previous studies of inter-event interferometry which focus on the relocation of earthquakes.\par

When dealing with coda waves, the scattering properties of the medium are of interest. Therefore with the help of the \emph{Qopen} software package \citep{Eulenfeld2016, Eulenfeld2017} scattering and intrinsic attenuation properties of shear waves were estimated for a homogeneous scattering medium in the half space.
The transport mean free path, defined by the average length in which the wave forgets its initial direction due to single or multiple scattering, was determined at approximately \SI{130}{km} with the data set used in this study for a frequency range between \SI{10}{Hz} to \SI{20}{Hz}. The corresponding transport mean free time is \SI{36}{s} for a shear wave velocity of \SI{3.6}{km/s}. The coda waves with a travel time up to \SI{50}{s} used in this study are therefore multiply scattered, but not in the diffusion regime. The intrinsic absorption length is estimated at \SI{55}{km} for a frequency range between \SI{10}{Hz} to \SI{20}{Hz}.
Similar values were obtained by \cite{Gaebler2015} and \cite{Bachura2016} for this region with different data sets.

\section{Data and method}

\label{sec:data}

\begin{figure}
\centering
\includegraphics[width=0.9\textwidth]{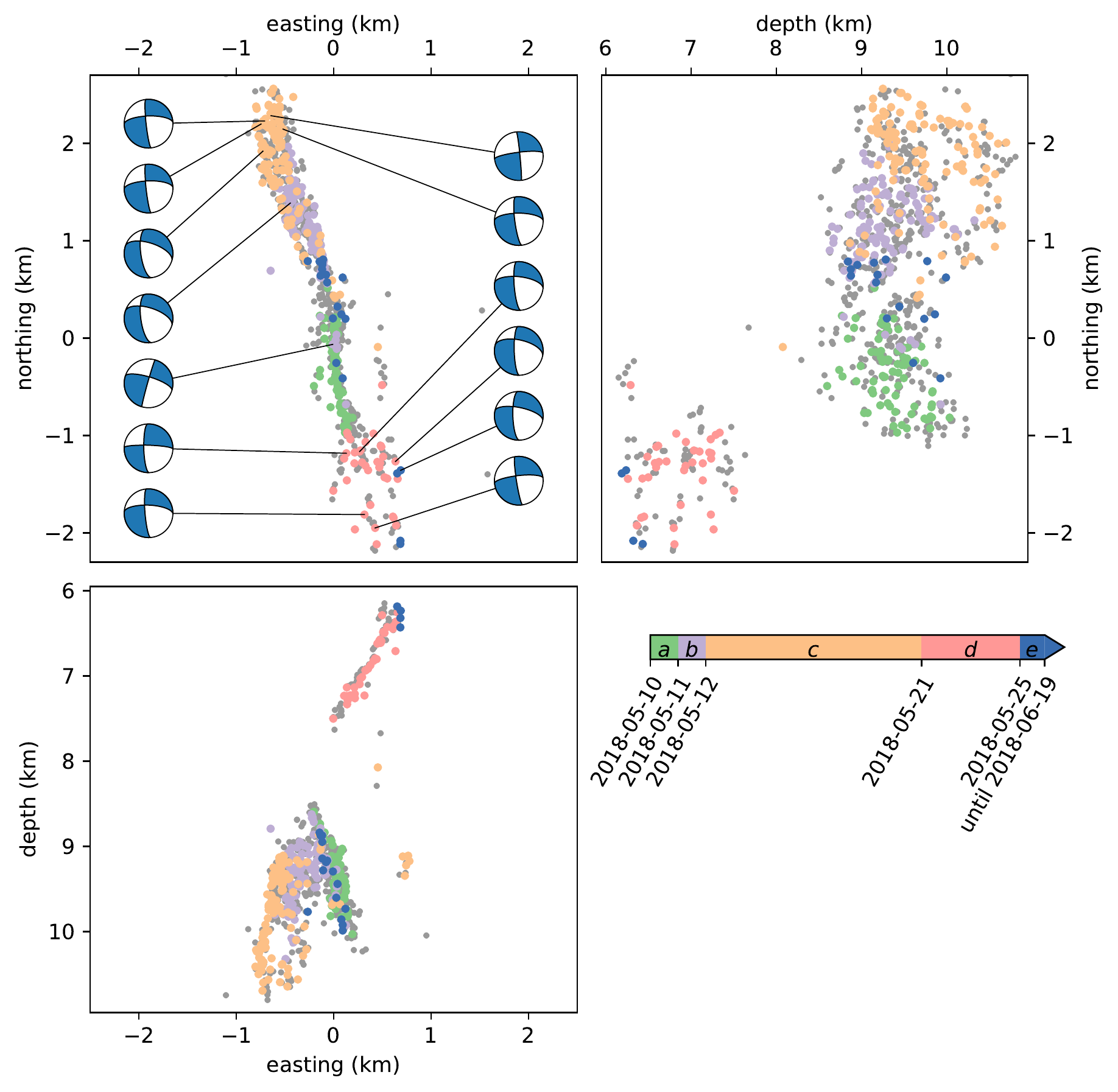}
\caption{Map and depth sections of all 966 events of the 2018 earthquake swarm \citep[gray,][]{Fischer2019AGU} with moment tensors of selected events \citep{Plenefisch2019EGU}. 376 events have a magnitude larger than 1.8 and are divided into 5 clusters \emph{a-e} with time (colored points).
See the main text for the exact definition of points in time separating the clusters. Coordinates in the map are relative to 50.25°N, 12.45°E.}
\label{fig:eventmap}
\end{figure}

The double difference catalog of the 2018 swarm used in this study was compiled by Martin Bachura \citep{Fischer2019AGU}. It consists of approximately 1000 earthquakes with local magnitudes larger than 1.3. The uncertainties in origin locations are \SI{50}m relative to each other.
For the purpose of the present study, the catalog is divided into 5 clusters named \emph{a-e} with earthquakes larger than magnitude 1.8 delimitated and separated by the times \mbox{2018-05-10}, \mbox{2018-05-11~03:10}, \mbox{2018-05-12~06:00}, \mbox{2018-05-21}, \mbox{2018-05-25}, \mbox{2018-06-19} (all in Coordinated Universal Time, UTC). The clusters are plotted in figure~\ref{fig:eventmap} in map view and depth sections. The small cluster south of the region indicated with the box in figure~\ref{fig:map} is excluded from the analysis.

Cluster~\emph a started the 2018 swarm activity at a depth of \SI 9{km} to \SI{10}{km}. In the following days, the activity migrated to the north (clusters \emph{a-c}). On May~21 earthquakes (cluster~\emph d) started rupturing a region south of the first cluster~\emph a at lower depth (around~\SI 7{km}). After June~19 the activity faded out with cluster~\emph e.
The event clusters are defined in time, but are with few exemptions equally separated in space as can be noticed from figure~\ref{fig:eventmap}.

Focal mechanisms of earthquakes larger than magnitude 3 are strike-slip mechanisms with a normal faulting component aligned with the Mariánské Lázně fault zone \citep{Plenefisch2019EGU}.
\par

The waveforms of the earthquakes were recorded on the Czech WEBNET stations with \SI{250}{Hz} sampling rate. For this study, data from 9 WEBNET stations are used (compare figure~\ref{fig:map}).

Waveforms are bandpass filtered in the frequency range \SI{10}{Hz} to \SI{40}{Hz}. Coda time windows are selected from \SI 1s after the S pick to \SI{50}s after the origin time. A shorter time window is taken if waves from another event in the catalog interfere. In this case, the coda window ends at the P pick of the following event. When the amplitude falls below a threshold based on the noise level the coda window ends at this time. All waveforms are visually inspected and the time window is adapted if necessary due to earthquakes not present in the catalog but interfering with the coda waves.

To compensate for intrinsic attenuation data is normalized by division by the instantaneous amplitude (i.e.\ envelope). This normalization step, later referred to as time normalization, ensures that data inside the coda window is equally weighted.
The following processing steps are applied to pairs of earthquakes. 
Pairs with an inter-event distance larger than \SI1{km} are excluded due to the low fraction of scattered waves which travel the direct path between the events.

Data of two different earthquakes recorded by the same station and channel are aligned relative to origin times. The overlap of the coda windows defines the time window $(t_1, t_2)$ used for the cross-correlation. Station-component combinations with time windows shorter than \SI{10}s are discarded. For each station and component the first signal $s_1$ from the shallower event and second signal $s_2$ from the deeper event are cross-correlated according to

\begin{equation}
C\fof t = \frac{
\displaystyle\int_{t_1}^{t_2} s_1\fof\tau s_2\fof{\tau - t} d\tau
}{
\left(\displaystyle\int_{t_1}^{t_2} s_1\fof\tau^2 d\tau\right)^{1/2} \left(\displaystyle\int_{t_1}^{t_2} s_2\fof\tau^2 d\tau\right)^{1/2}
}\,.
\label{eq:cc}
\end{equation}

\begin{figure}
\centering
\includegraphics[width=0.9\textwidth]{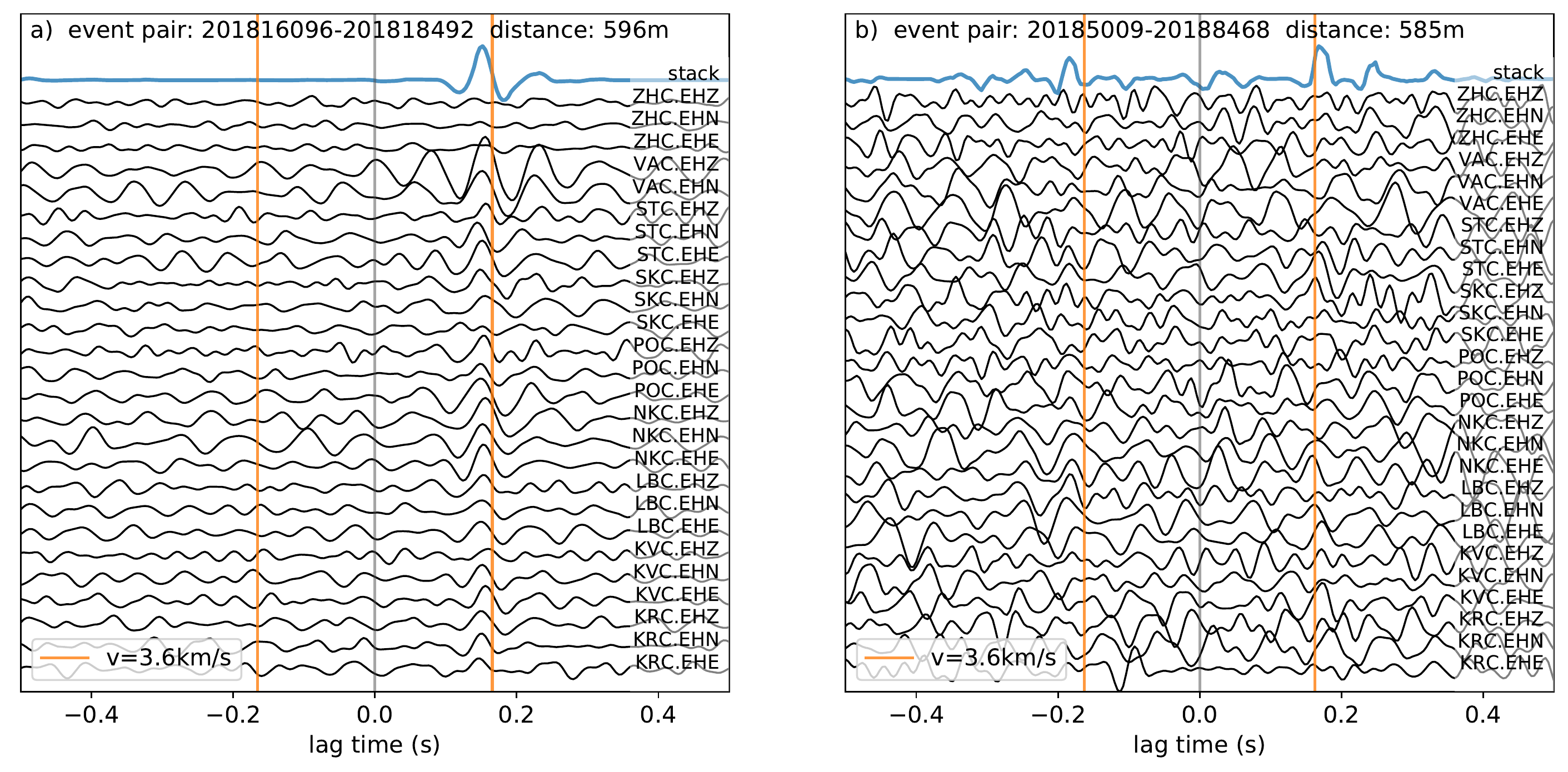}
\caption{Cross-correlation functions for two different event pairs with phase weighted stack at the top. The event pair in the left panel~(a) shows peaks in the cross-correlation function on most stations and components at a lag time which corresponds to a direct wave traveling between earthquake locations with a velocity of \SI{3.6}{km/s} (orange lines). Contrary, for the event pair in the right panel~(b), peaks at expected lag times only emerge in the phase weighted stack.}
\label{fig:cc}
\vspace*{\floatsep}
\centering
\includegraphics[width=0.9\textwidth]{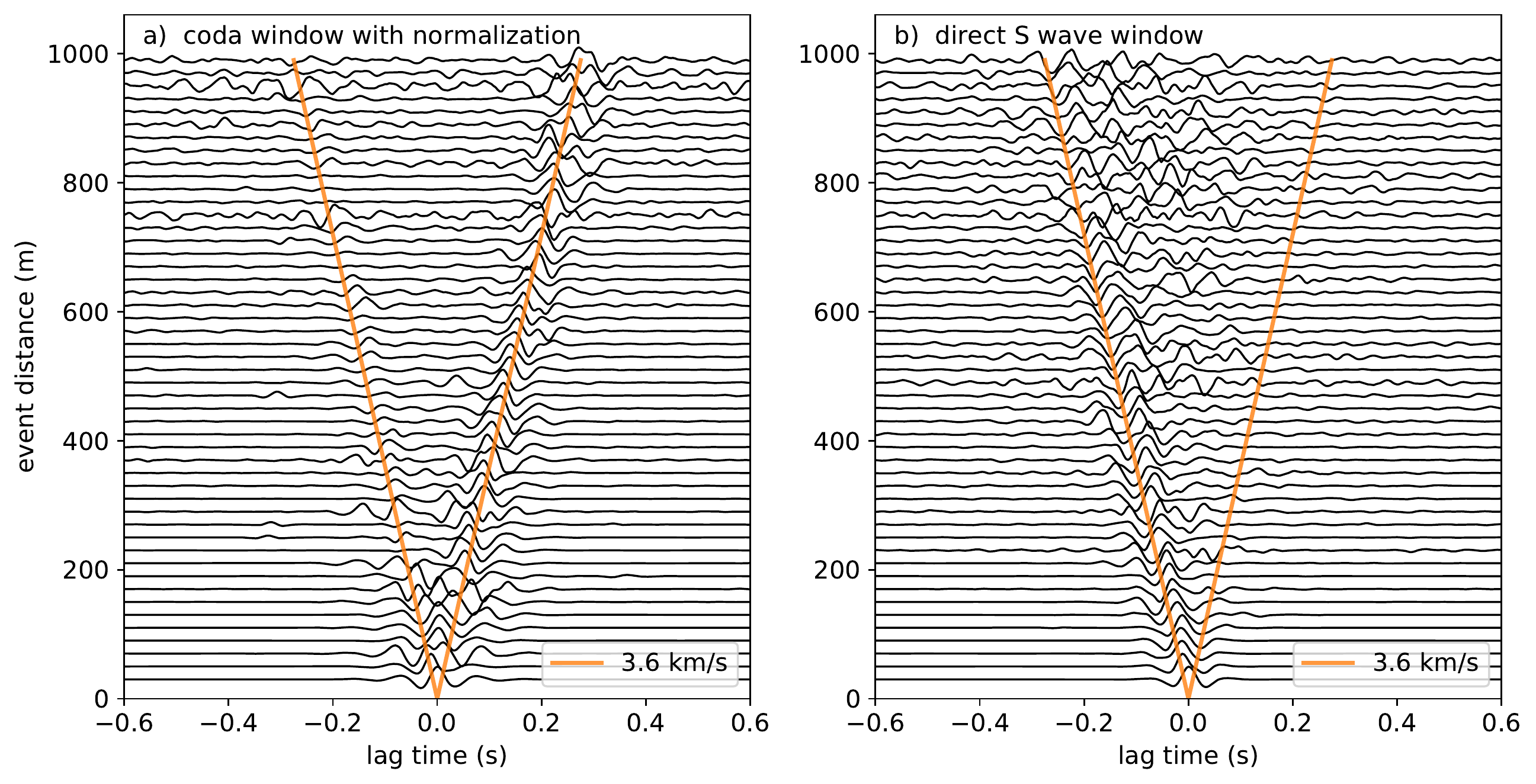}
\caption{Cross-correlation functions stacked versus distance between earthquakes. In the left panel~(a) cross-correlations are calculated for time-normalized data in the coda time window (method~C in figure~\ref{fig:methods}). For comparison, results for the direct S wave window without time-normalization (method~B) are displayed in the right panel~(b).
A clear move-out with a velocity of around \SI{3.6}{km/s} is visible for the coda window.
For the direct S wave window, the apparent travel time associated with random peaks in the cross-correlation function is lower than the expected inter-event travel time.}
\label{fig:cc_vs_dist}
\end{figure}

Only correlations between the components Z-Z, N-N and E-E are used; cross-component correlations did not show a peak with a similar height at the expected time interval.

For each event pair all cross-correlation functions for different stations and components are stacked together. To enhance the signal a phase-weighted stack of order 2 is used \citep{Schimmel1997} and in the process each trace is weighted by the length of the coda window used for the cross-correlation. Figure~\ref{fig:cc} displays examples of cross-correlations for two different event pairs. In the left panel~\ref{fig:cc}a peaks corresponding to waves traveling from the location of one earthquake to the location of the other earthquake can be identified on each individual trace. Contrary, in the right panel~\ref{fig:cc}b the peaks corresponding to the inter-event travel time only emerge after the stacking procedure.

The relationship between the lag time of the maxima \tmax resulting from waves traveling between the earthquake locations and the inter-event travel time \ttr is

\begin{equation}
\tmax = \ttr\sgn\fof{\tmax} + \dto 1 - \dto 2 + \delta\ind{err}\,.
\label{eq:tmax}
\end{equation}

$\sgn$ is the sign function and \dto 1 and \dto 2 are the errors of the origin times of the first and second earthquake, i.e.\ the difference between catalog origin time and real origin time.
If spatial deviations in seismic heterogeneity are present around the two sources, the lag time of the maxima might slightly deviate from the travel time corresponding to the stationary point by the error $\delta\ind{err}$. Because of the high dominant frequencies used in this study, a relatively homogeneous scatterer distribution is expected in the stationary phase zone and this error is therefore neglected in the following.

When two peaks \tmaxi 1 and \tmaxi 2 corresponding to waves traveling between event locations in both directions are visible in the cross-correlation function as in figure~\ref{fig:cc}b, the inter-event travel time and difference in origin time errors can be determined directly with

\begin{align}
\ttr & =  \frac 12 \left|\tmaxi 1 - \tmaxi 2\right|\,,\\
\dto 1 - \dto 2 & = \frac 12 \parent{\tmaxi 1 + \tmaxi 2} \,.
\label{eq:ttr}
\end{align}

Equation~\ref{eq:ttr} can be used to check the quality of the earthquake catalog.
In the scope of this study, errors of origin times are ignored and the inter-event travel time is approximated by the lag time of a single maximum: $\ttr \approx \left|\tmax\right|$.

In figure~\ref{fig:cc_vs_dist}a the phase weighted stacks of coda cross-correlations of each event pair are linearly stacked in different bins of event distance and plotted versus event distance. A clear move-out with a velocity of around \SI{3.6}{km/s} indicates that the peaks in the cross-correlation functions are due to shear waves traveling between the earthquake locations and that the errors in origin time are much smaller than inter-event travel times.

For comparison, the same procedure is repeated for the direct S wave window $(\SI 0s, \SI 2s)$ relative to the S pick without time normalization (method~B).
When using the direct waves, spurious arrivals in the cross-correlation functions only cancel out if the station distribution is sufficiently dense. This corresponds to the requirement of a sufficiently random wave field in the inter-receiver setting.

The usual procedure to eliminate spurious arrivals in method~B is the exclusive selection of stations inside the stationary phase zone, but because of the governing geometry the stationary phase zone is relatively small at the surface and no stations or only a small amount of stations may be located within the zone. Therefore, cross-correlations at all stations and components are simply aggregated with a phase weighted stack of order 2. The linear stacks in bins of event distance are displayed in figure~\ref{fig:cc_vs_dist}b.
Not a single peak, but rather a multitude of peaks at absolute lag times smaller than the expected inter-event travel time emerge.
This expected result limits the usability of method~B compared to method~C for the present application.

\section{Results}

\begin{figure}
\centering
\includegraphics[width=0.9\textwidth]{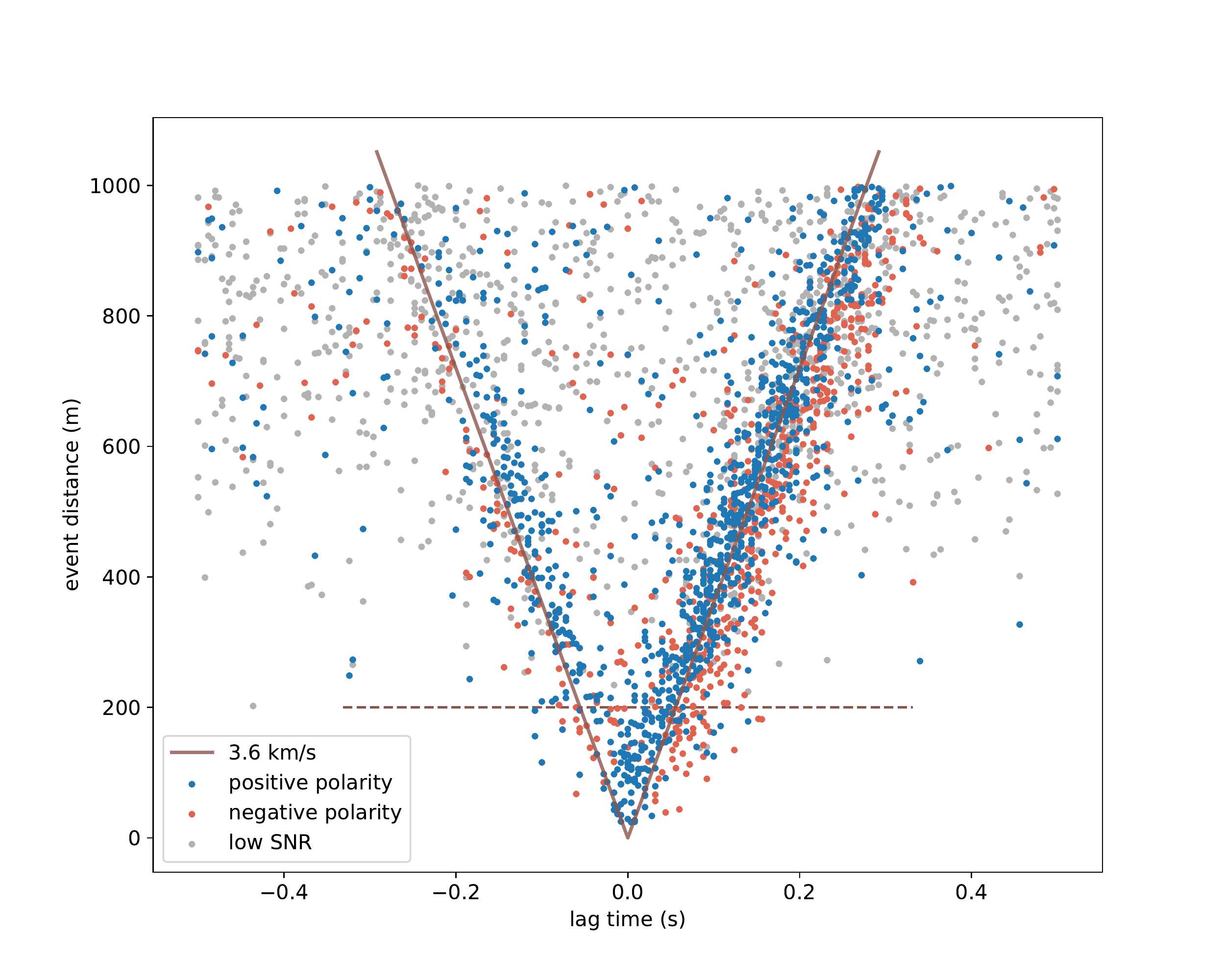}
\caption{Lag times of maxima of cross-correlation functions for each earthquake pair versus distance between the two earthquakes. The polarity of the peaks is indicated in blue and red, peaks with low signal-to-noise ratio are plotted in gray. Most data points scatter around the move-out for a velocity of \SI{3.6}{km/s}. The dashed horizontal line at \SI{200}m marks the wavelength for a frequency of \SI{18}{Hz} and velocity \SI{3.6}{km/s}.
}
\label{fig:max_vs_dist}
\end{figure}


Maxima with positive or negative polarity are extracted from the phase weighted stack for each event pair. To avoid the definition of a noise time window, the noise level is arbitrarily defined as the absolute value of the 8th largest relative extremum in the stacked cross-correlation function. Figure~\ref{fig:max_vs_dist} displays the lag times of the maxima with positive or negative polarity for all event pairs. Several interesting features can be observed in this figure.

First of all, the data points resemble a move-out of around \SI{3.6}{km/s}, although a considerable amount is off target. Therefore, around 1850 of 2900 data points are selected for this study by enforcing a signal-to-noise ratio larger than 10. Due to this condition a large amount of outlying data points are discarded (gray points in figure~\ref{fig:max_vs_dist}). The observed move-out shows that most of the peaks arise due to the extraction of shear waves traveling between the earthquake locations. Note that, maxima corresponding to inter-event travel times of compressional waves are not observed.
This indicates that initial P waves are less likely be recorded in the S wave coda than initial S waves which is consistent with the high energy ratio of radiated S to P  waves for earthquakes (see section~\ref{sec:intro}).
The maxima therefore reflect the inter-event travel times of shear waves.

This is not the case for small event distances for which the maxima in the cross-correlation function are located at smaller lag times than expected from the distance (also compare with figure~\ref{fig:cc_vs_dist}a). Then, coda waves from both events approximately travel along the same paths for all radiation directions and the event pairs are suited for a decorrelation analysis (method~A).
This is the near field that is also excluded in inter-receiver interferometry.
A typical wavelength for the data set is \SI{200}m (frequency \SI{18}{Hz}, velocity \SI{3.6}{km/s}) and marked as a horizontal line in figure~\ref{fig:max_vs_dist}. Event pairs with a shorter distance are excluded from further analysis.
\par

\begin{figure}
\centering
\hfill
\begin{minipage}{0.3\textwidth}
\includegraphics[width=\textwidth]{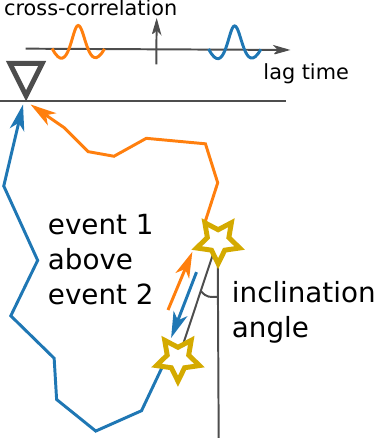}
\end{minipage}
\hspace{0.05\textwidth}
\begin{minipage}{0.55\textwidth}
\includegraphics[width=\textwidth]{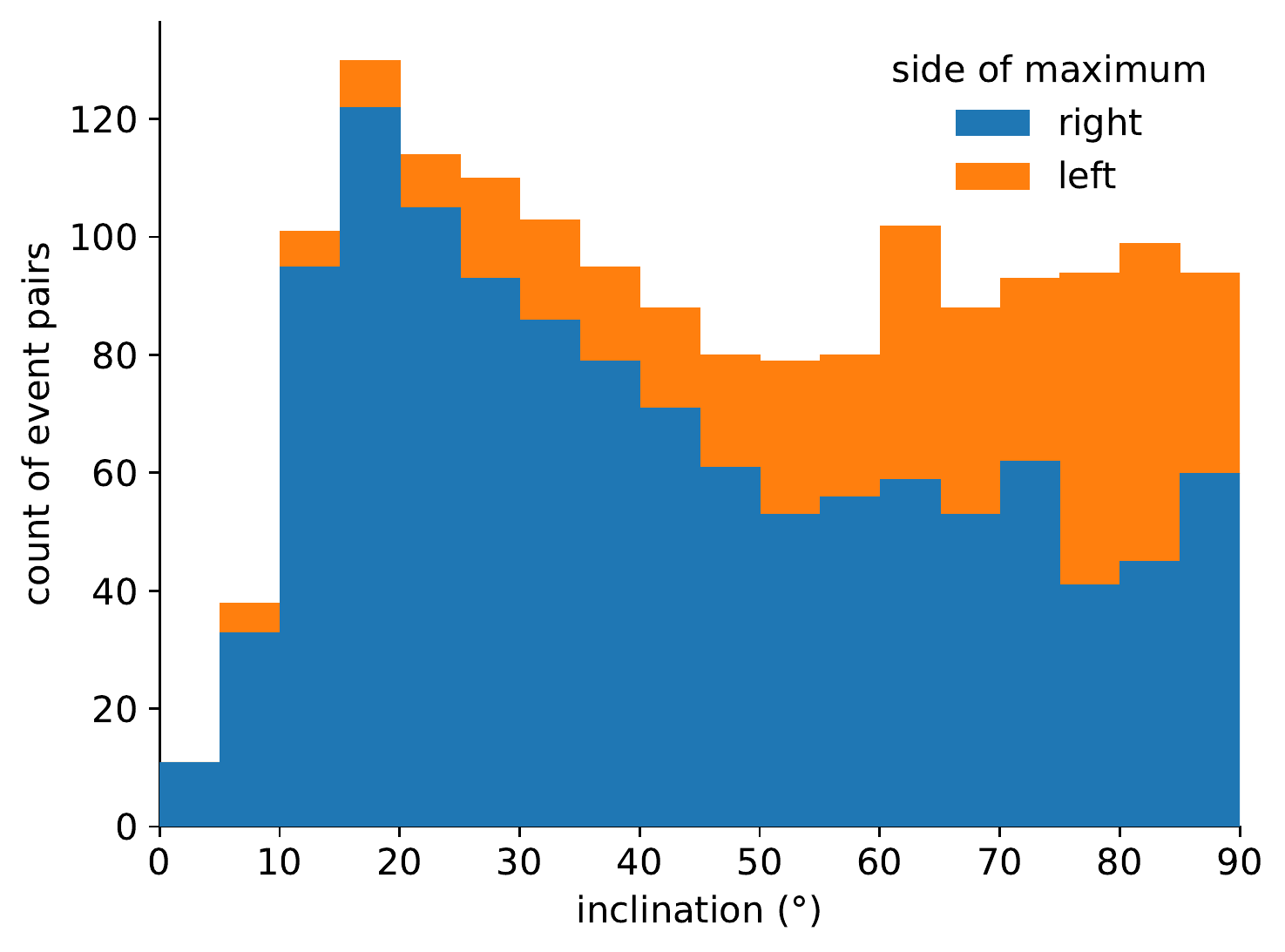}
\end{minipage}
\hfill
\caption{Side of maxima in cross-correlation function depending on inclination between the locations of the two used earthquakes. The waveform of the shallower earthquake is used as first signal in the cross-correlation. Maxima in the right side of the cross-correlation function correspond to waves traveling from the first to the second event; maxima in the left side vice versa (compare sketch).
}
\label{fig:hist}
\end{figure}

The second observation is, that considerably more data points are located on the right side than on the left side of the cross-correlation function. Because the waveform of the shallower event is used as the first signal in the cross-correlation calculated with equation~\ref{eq:cc},
maxima on the right side (positive side) of the cross-correlation function correspond to waves traveling from the shallower to the deeper event; maxima on the left side (negative side) vice versa.
Figure~\ref{fig:hist} shows a stacked histogram of the number of data points on the right and left side of the cross-correlation function as a function of inclination angle between the two events. For events with a low inclination angle  (different depth, similar epicenter) maxima on the right side of the cross-correlation function are predominant indicating that waves going upwards are less likely to be recorded as coda than waves which leave the source in the downward direction.

A higher intrinsic attenuation in the top \SI{10}{km} could explain this observation, because coda waves that are initially going upwards spend more time in the uppermost crust and would suffer stronger attenuation than the initially downgoing waves.

For high inclination angles, events are located at similar depth and maxima on both sides of the cross-correlation are observed with similar frequency.\par

\begin{figure}
\centering
\includegraphics[width=0.9\textwidth]{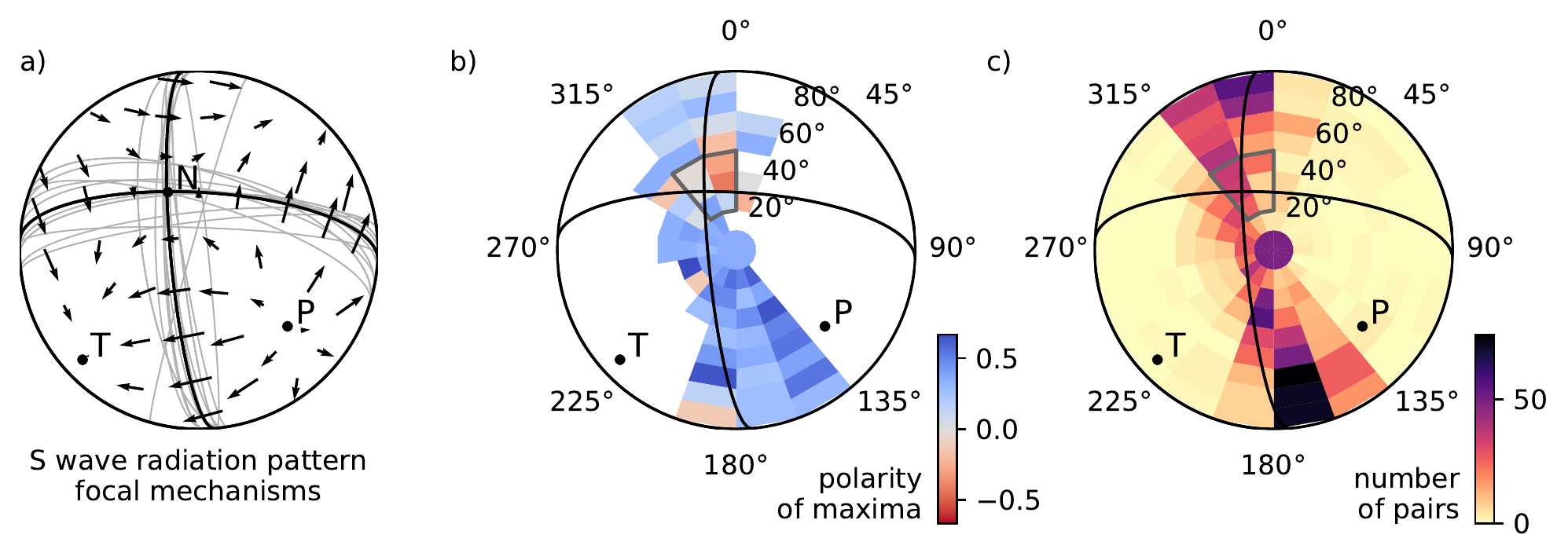}
\caption{a) Focal mechanisms of selected earthquakes (gray) and the median focal mechanism (black) with axis of largest compression and dilatation and neutral axis (P-, T-, N-axis). The corresponding S wave radiation pattern \citep[arrows,][chapter~4.3]{Aki2002} has zero points at the P-, T- and N-axis. The largest amplitudes are expected at fault plane and auxiliary plane.\newline
The median focal mechanism with its P- and T-axis is also displayed in the two other panels together with a binned polar plot of the average of polarities of maxima of cross-correlation functions (b) and with a polar histogram of the number of event pairs (c). Negative polarity occurs preferentially for specific orientations between the two earthquakes and is near the N-axis. Because of unstable polarity and low amplitude of radiated shear waves near the N-axis, event pairs with azimuth and inclination inside the region marked by the gray line are later excluded from analysis. All polar plots use a lower-hemisphere stereographic projection with labels of inclination and azimuth values. In the middle panel only bins with more than 5 event pairs are displayed.
}
\label{fig:focal}
\end{figure}

Thirdly, maxima of cross-correlation functions predominantly have a positive polarity. Maxima with negative polarity can be observed; often these have a larger time lag and arise after positive maxima of event pairs with similar distance (compare figure~\ref{fig:max_vs_dist}).
This observation could indicate that the correct inter-event travel time is associated with the positive maximum.
However, due to its oscillatory character the correlation function shows a down swing following the positive peak which might in some cases be of larger amplitude due to noise. 
To measure the wave velocity it is therefore desirable to search only for maxima with positive polarity.
Before, cases in which a truly negative peak can occur at the correct time have to be identified and excluded.\par

Different focal mechanisms of the two events in the pair can lead to a negative peak, because the polarity of radiated S waves into the inter-event direction might be different for both events (figure~\ref{fig:focal}).
Simulations performed by \cite{Sun2016} confirm this insight.

Figure~\ref{fig:focal}a displays the focal mechanism of 13 larger earthquakes determined by \cite{Plenefisch2019EGU}. The variability in these 13 mechanisms gives an idea of the variability of the focal mechanisms of all earthquakes in the swarm under the reasonable assumption that most smaller earthquakes have a similar focal mechanism.

The average of the observed polarity of the maxima in the cross-correlation function and the number of event pairs are displayed in figure~\ref{fig:focal}b and \ref{fig:focal}c in binned polar plots as a function of inter-event inclination and azimuth. Note, that near the N-axis both positive and negative polarities can be observed. For other directions positive polarity is predominant. The distribution of number of event pairs reflects the shape of the 2018 earthquake swarm.

For the median focal mechanism with median slip, rake and dip the S wave radiation pattern according to chapter 4.3 of \cite{Aki2002} is plotted alongside. No energy is radiated along the axis of largest compression and dilatation and neutral axis (P-, T-, N-axis). The maximal energy of shear waves is radiated along the fault plane and auxiliary plane. 

For exactly the same focal mechanism shear waves of the same polarity are emphasized by the cross-correlation function independent of the proximity of the two events and a peak of positive polarity will emerge in the cross-correlation function. For similar focal mechanisms a peak of positive polarity is still expected for all radiation directions except near the P-, T- and N-axis, where the shear wave polarity may be completely different for both events. Additionally, the amplitude of the radiation in inter-event direction is diminutive near the P-, T- and N-axis.

Therefore, depending on the exact orientation and focal mechanisms of the two earthquakes a peak of positive or negative polarity can arise in the cross-correlation function at the inter-event travel time or there may be no peak at all at this lag time \citep{Sun2016}. 

In order to circumvent these error-prone situations and in order to exclude a possible negative polarity at the correct time, 
event pairs with azimuths larger than 320° and inclinations between 20° and 50° (region marked in figure~\ref{fig:focal}), all near the N-axis of the median focal mechanism, are excluded from the further analysis.
\par

\begin{figure}
\centering
\includegraphics[width=\textwidth]{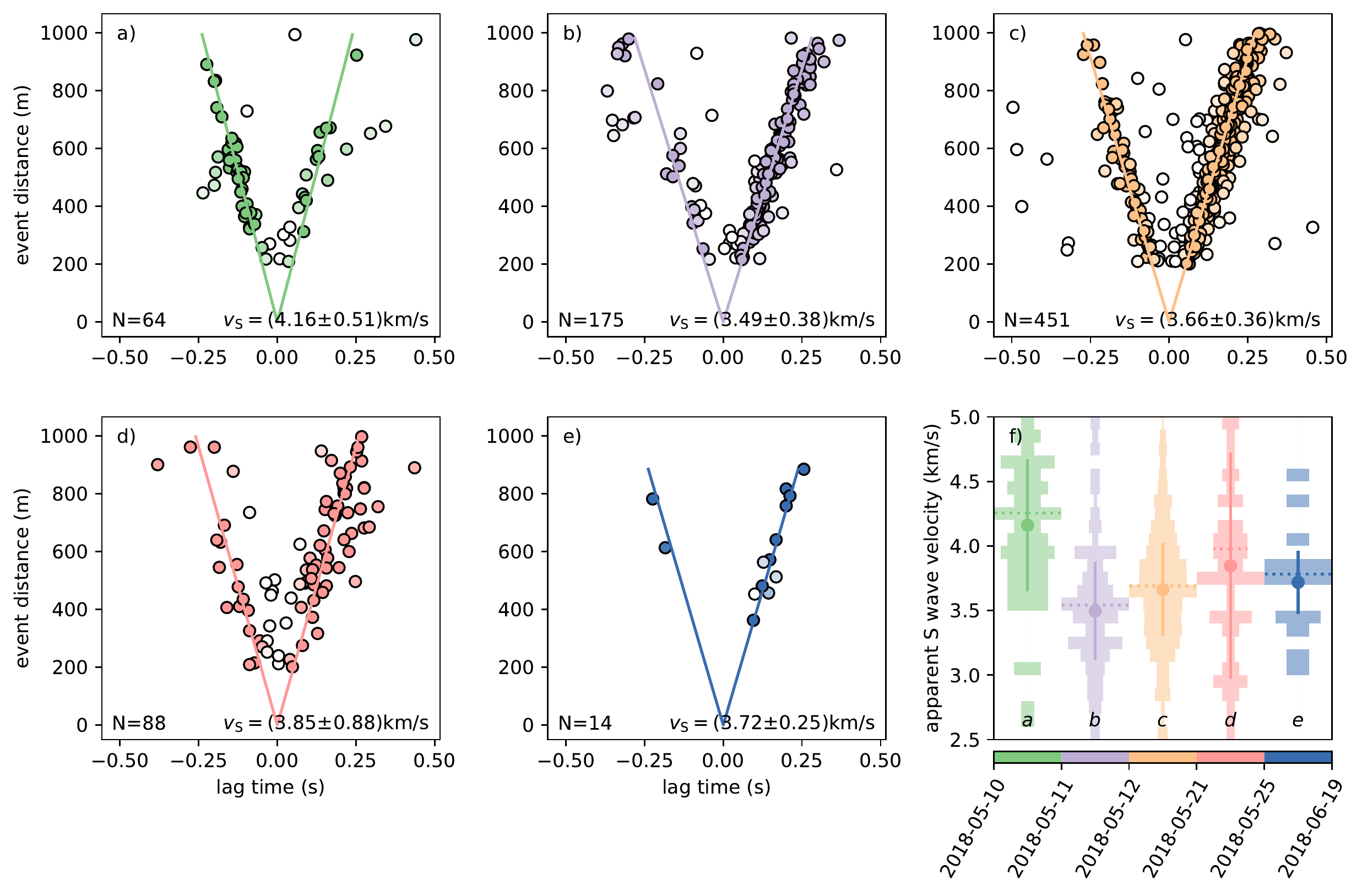}
\caption{a-e) Lag times of maxima of cross-correlation functions for event pairs of different earthquakes clusters \emph{a-e}. See figure~\ref{fig:eventmap} for a map of the different clusters and the main text for the points in time defining the clusters. The line in each panel corresponds to the robustly determined mean velocity in each cluster. The opacity of each data point is proportional to its weight in this robust inversion. Each panel also indicates the number of event pairs and the value of the robust mean with its median absolute deviation (MAD).
f) In the lower right panel the velocity for each cluster is plotted with its MAD as errorbar over a histogram of apparent velocities of each event pair. Additionally the median apparent velocity is marked with a horizontal dotted line. The velocity of the first cluster~\emph a is significantly larger than the velocities in the other clusters \emph{b-e}.
}
\label{fig:vs_time}
\end{figure}

Finally, figure~\ref{fig:vs_time}a-e displays the lag times of positive maxima of the cross-correlations functions over event distance for the different event clusters \emph{a-e}. Beside the previously mentioned constraints both events need to be from the same cluster. The velocity inside each cluster is determined by a robust mean \citep{Eulenfeld2016, Huber2014} of the apparent velocities of each event pair. The weighting function for the iteratively reweighted least squares (IRLS) algorithm is the Hampel's 17A function with tuning constants $a{=}1$, $b{=}2$ and $c{=}3$. Panel~f of figure~\ref{fig:vs_time} displays the robustly determined shear wave velocities inside the five clusters together with its median absolute deviation (MAD) as error. The distribution of apparent seismic velocities in each event cluster is also indicated, together with its median value. The determined values for the shear wave velocities together with the corresponding MADs are given in panels~a-e of figure~\ref{fig:vs_time}. The velocities range from \SI{3.49}{km/s} to \SI{4.16}{km/s} with MADs between \SI{0.25}{km/s} and \SI{0.88}{km/s}. Cluster \emph a has a significantly larger velocity of \SI{4.16(51)}{km/s}
and cluster~\emph d a slightly larger velocity of \SI{3.85(88)}{km/s} than the other clusters \emph b, \emph c and \emph e with velocities between \SI{3.49}{km/s} and \SI{3.72}{km/s}. The variation around the mean velocities is high, especially for cluster~\emph d.

\section{Discussion}

\subsubsection*{Comparison between different methods of inter-source interferometry}

All of the three methods introduced in section~\ref{sec:intro} aim at extracting the travel times between event locations. Method~C was used in this study. Method~B and C make use of event pairs which are separated by a larger distance than the dominant wavelength compared to method~A for which event pairs must be separated by less than the dominant wavelength.
More event pairs fulfill this requirement for method~B and C and therefore more data can be collected than for method~A.
Method~B which uses the direct wave train and method~A do have the advantage that the waveforms between the event pair are expected to be more similar compared to method~C. This makes outlying data points as in figure~\ref{fig:max_vs_dist} less likely.
A problem with method~B is the constraint that stations have to be present in, but also be restricted to the stationary phase zone. Depending on the data set this can lead to a very low amount of stations or even no stations which can be used in the analysis; this was the case for the present data set.

The assumptions for method~A are stronger than for method~C. To calculate the event separation from the decorrelation value the same focal mechanism for the two earthquakes and the same scattering and attenuation strength in all directions is required \citep{Snieder2005}. The second assumption is not valid for the present application, otherwise a balanced distribution of maxima between the left and right side of the cross-correlation function would be expected (compare figure~\ref{fig:hist}).

Method~C in contrary assumes 1) a heterogeneous scatterer distribution in the stationary phase zone and 2)
\emph{similar} focal mechanisms to the extent that the polarity of waves which are radiated into the inter-event direction is the same for both events. The same polarity can be guaranteed for similar focal mechanisms by excluding pairs with inter-event directions near the P-, N- or T-axis (compare figure~\ref{fig:focal}). 
This step and the underlying constraint of similar focal mechanisms could also be omitted in favor of a more dedicated quality check of the calculated cross-correlations and their maxima or by taking into account the focal mechanisms of the earthquakes and their impact on the polarity of maxima in the inter-event cross correlation functions.

Processing for method~A is sophisticated compared to the straight-forward processing of method~C.
On the other hand, method~A has the advantage over method~C to not depend on the time lag of the maxima in the cross-correlation function. Therefore a possible error in origin times does not play a role as in method~C.

\subsubsection*{Shear wave velocities for different earthquake clusters}

The inter-event interferometry was for the first time applied to map shear wave velocity in the source region of a fluid-driven earthquake swarm. 
The measured shear wave velocity $v\ind S{=}\SI{4.16(51)}{km/s}$ of the first cluster~\emph a is significantly higher than the velocities of the clusters~\emph b, \emph c and \emph e of around 3.49 to \SI{3.72}{km/s}. The measured velocity for cluster~\emph d is around \SI{3.85}{km/s} and higher than the mean velocity of approximately \SI{3.6}{km/s} for the three clusters \emph b, \emph c and \emph e. The data points of cluster~\emph d have a high median absolute deviation of \SI{0.88}{km/s} which could indicate that this cluster contains subclusters with different apparent velocities.
The higher velocities of clusters~\emph a and \emph d are robust, insofar they are consistent if some of the applied constraints are relaxed or changed.

The method as applied in this study assumes that the used earthquake catalog is correct in origin times and locations.
An error in origin times shifts the determined inter-event travel time by the difference of the error in origin time of the first and second event (see equation~\ref{eq:tmax}).
For errors in origin locations the distance between the events is changed.
Both errors affect the velocity estimate and it is expected that the high variability of apparent velocities of different event pairs in one cluster is due to them. However, these errors do not introduce a systematic bias, but lead to a higher median absolute deviation.

For the future I suggest to extent the presented method to a full-fledged tomography by taking into account direct wave onsets, double difference travel times \citep[TomoDD,][]{Zhang2003} and inter-event travel times and by inverting for origin times, locations and a seismic velocity grid in a single joint inversion.

\subsubsection*{Interpretation of variance in shear wave velocity}

\begin{figure}
\centering
\includegraphics[width=\textwidth]{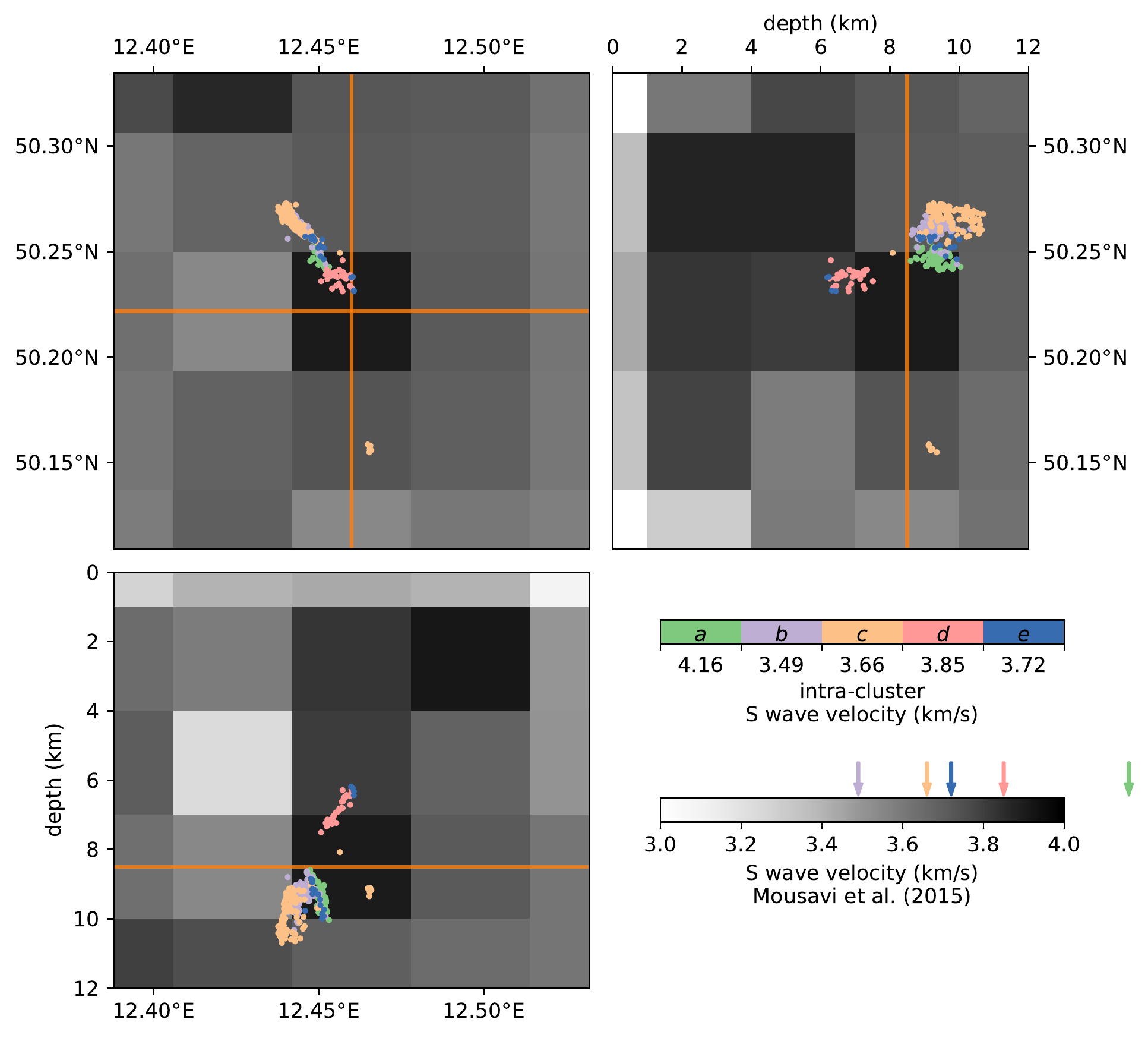}
\caption{Map and depth sections of the shear wave velocity model of \cite{Mousavi2015} together with the event clusters defined in section~\ref{sec:data}.
The map is a horizontal slice of the model at depth \SI{8.5}{km} (orange lines in depth sections). The depth sections are vertical slices at 12.46°E longitude and 50.22°N latitude (orange lines in map). The clusters~\emph a and \emph d for which high velocities were obtained (green color and light red color) are located inside a model box with high shear wave velocity.}
\label{fig:mousavi}
\end{figure}

Because there is neither a temporal nor spatial overlap between the defined earthquake clusters it is difficult to tell from the results alone if the differences in shear wave velocity are of temporal or spatial nature. Therefore, I will compare the results to \cite{Dahm2013} who observed a temporal variability and \cite{Mousavi2015} who performed a classical travel time tomography with spatial information.

\cite{Dahm2013} observed P~wave to S~wave velocity ratios ({\vpvsratio}s) lower than 1.45 at the beginning of the swarms in the years 1997, 2000 and 2008. Such low values of {\vpvsratio} have been observed earlier for the fault region of 1997 and 2008 swarms \citep{Vavrycuk2011}.
\cite{Bachura2016a} again observed lower velocity ratios down to 1.59 for some clusters of the 2014 earthquake swarm.
\cite{Dahm2013} use the Gassmann equations \citep[chappter~6.3]{Mavko2009} which describe seismic velocities for porous fluid saturated rocks as a function of porosity. For these equations the shear wave velocity does \emph{not} depend on porosity. \cite{Dahm2013} argue that the reduced velocity ratio is caused by a transition of the pore fluid to the gaseous phase and a resulting decrease in P~wave velocity. 

A temporary decrease in P~wave velocity could result in a systematic error of the underlying double-difference earthquake catalog at that period of time which in turn might lead to a higher observed shear wave velocity as in this study.
It is therefore important to calculate and examine the {\vpvsratio}s of the five defined clusters of the 2018 swarm. In the appendix~\ref{sec:vpvs} I show that the velocity ratio of the five clusters is approximately fixed at 1.68 (figure~\ref{fig:vpvs}) which is a typical value for this region \citep[e.g.][]{Malek2005}. Please note that no assertion is made about a possible reduction of \vpvsratio at the beginning of the individual clusters. The fixed \vpvsratio indicates that the observed higher shear wave velocity of the first cluster~\emph a is not due to a temporal variability of $\vp{/}\vs$, but rather a spatial variation in shear wave velocity.

\cite{Mousavi2017} speculate that a high attenuating zone ascertained at the same location as the velocity anomaly of this study, could be an indication of uprising fluids in permeable channels.

With a \vpvsratio of 1.68 the observed S wave velocity translates into a P wave velocity of \SI{7.0}{km/s} for cluster~\emph a and P wave velocities down to approximately \SI{6.0}{km/s} for the swarm stages \emph b, \emph c and \emph e. This value is consistent with the CEL09 seismic profile at depth of around \SI{10}{km} \citep{Hrubcova2005}. In figure~\ref{fig:mousavi} the event clusters are plotted into horizontal and vertical slices of the shear wave model of \cite{Mousavi2015}. 
As a matter of fact, the patch ruptured by cluster~\emph a with highest observed velocity is located in a model box with high shear wave velocity. Cluster~\emph d which also ruptured a region with increased velocity is located partly in the same box meaning that the higher shear wave velocity is consistent with previous observations.
The resolution of the model of \cite{Mousavi2015} is bounded on the low end by its block size of $\SI 4{km}{\times}\SI 4{km}{\times}\SI 3{km}$. The resolution in this study is about the extent of the earthquake clusters (${\approx}\SI 1{km}$) and already better than in a classical tomography of the same region. The resolution in a tomographic study based on inter-event interferometry depends on the inter-event distances (${<}\SI{100}m$). Therefore, the resolution is expected to be superior inside the swarm regions compared to classical tomography once the results of this study are used for a tomographic inversion.



\section{Conclusions}

A method of coda wave inter-source interferometry was applied to determine the shear wave velocity in different parts of the source volume of the 2018 West Bohemia earthquake swarm. Although, similar methods were used before for earthquake relocation procedures, this is the first study that allowed to map the seismic velocity. This is facilitated by a high quality earthquake catalog that has been relocated with the double difference technique. The work should be considered as a preparatory study for source region tomography and therefore several features visible in the inter-event cross-correlation functions were explained.
The resolution of seismic velocity of this study is already better than classical tomography studies of the same region.  I expect the resolution to improve further when using insights of this study within a joint tomographic approach.

\paragraph{Acknowledgments}
\begin{footnotesize}
I thank Martin Bachura for providing data from WEBNET seismic stations \citepalias{WebnetData} and for providing the double difference earthquake catalog. Furthermore I thank Laura Barth and Thomas Plenefisch for the focal mechanisms.
All data used in this study can be downloaded at \url{https://doi.org/10.5281/zenodo.3741465} \citep{dataset_swarm2018}.
Discussions with Christoph Sens-Schönfelder culminating in a full-blown review of the manuscript are very much appreciated.
I thank Jenny Borns and Ulrich Wegler for proof-reading the manuscript.
Comments from two anonymous reviewers and associate editor Nori Nakata considerably improved the manuscript.
Data processing and plotting was performed with ObsPy \citep{Megies2011} and the Python packages NumPy, SciPy, matplotlib and statsmodels \citep{NumPy, SciPy, Hunter2007, statsmodels}.
This article can be reproduced with the source code provided at \url{https://github.com/trichter/inter_source_interferometry} \citep{Eulenfeld2020_sourcecode}.
\end{footnotesize}

\appendix

\section{Source region velocity ratios}
\label{sec:vpvs}
\begin{figure}
\centering
\includegraphics[width=\textwidth]{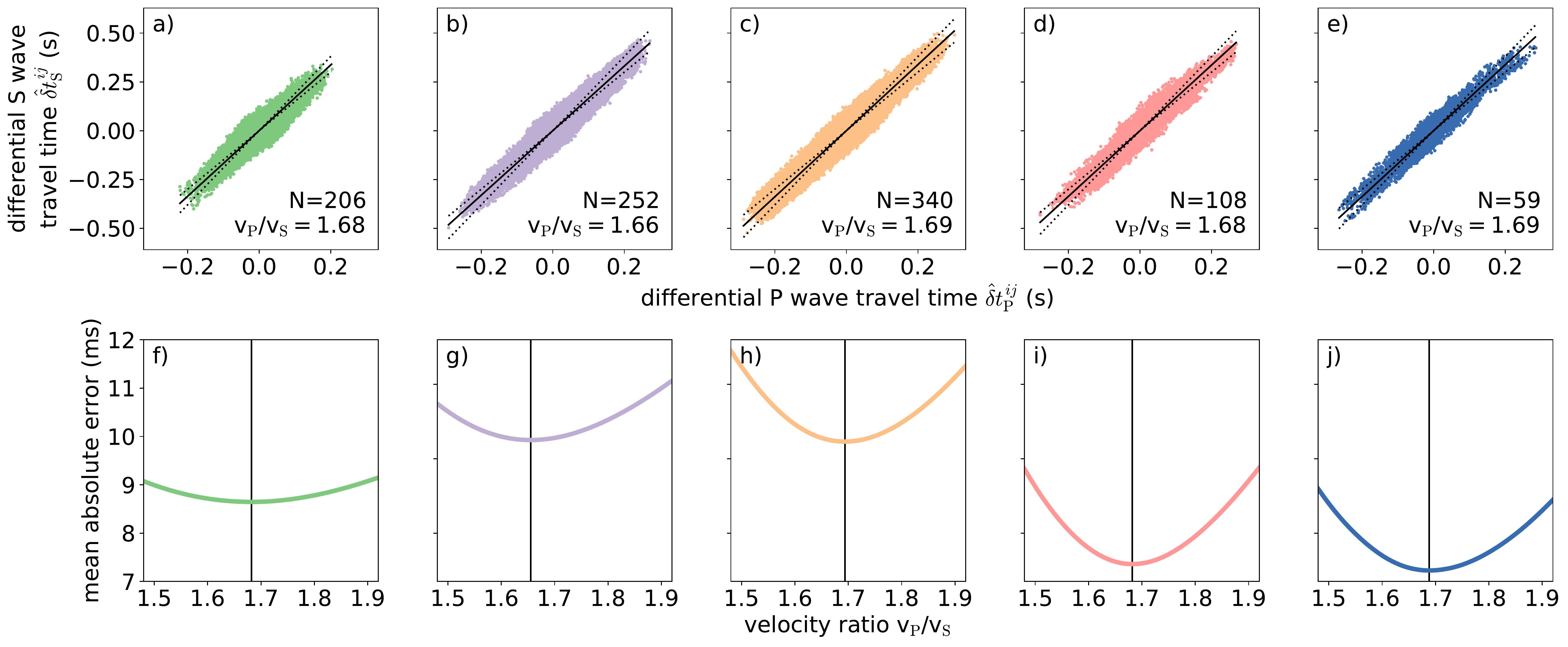}
\caption{a-e) Velocity ratios $v\ind P/v\ind S$ for each earthquake cluster \emph{a-e} from double difference travel times. Displayed is the orthogonal distance regression with L1 norm (straight line) and velocity ratios of 1.5 and 1.9 (dotted lines). The number of used events and the determined velocity ratio is displayed in each panel.
f-j) For each cluster \emph{a-e}, the mean absolute error (orthogonal distance) is displayed for tested velocity ratios.}
\label{fig:vpvs}
\end{figure}

To check whether the observed variability of shear wave velocity involve changes in the \vpvsratio, the method of \cite{Lin2007} is applied to the data set of this study.
The method can be explained with the following equations

\newcommand{\dt}[1]{\ensuremath{\delta t\ind{#1}^{ij}}\xspace}
\newcommand{\dth}[1]{\ensuremath{\hat{\delta}t\ind{#1}^{ij}}\xspace}
\newcommand{\dtm}[1]{\ensuremath{{<}\delta t\ind{#1}^{ij}{>}_j}\xspace}

\begin{align}
\frac{\dth S}R & = \frac 1R \frac{v\ind P}{v\ind S} \dth P \label{eq:vpvs1} \,\,\text{ and}\\
\dth P & = \dt P - \dtm P \,, \hspace{0.5cm}
\dth S  = \dt S - \dtm S \,.
\label{eq:vpvs2}
\end{align}

\dt P and \dt S are the travel time difference of P resp. S waves for an event pair with index $i$ at station with index $j$.
Because all event pairs are inverted together, the means over all stations \dtm P resp.\ \dtm S have to be subtracted from data points of each event pair. The \vpvsratio can be obtained by a regression between differential S wave travel times \dth S and differential P wave travel times \dth P. $R$ is a scaling factor, which is initially set to 1.

For the present data, picks from all available earthquakes inside each event cluster (including earthquakes with a magnitude lower than 1.9) are used to calculate \dt P and \dt S. In the first step, not the mean, but the median is used to calculate preliminary differential travel times \dth{P, prel}, \dth{S, prel} (equation~\ref{eq:vpvs2}).

Erroneous data points with
\begin{equation}
\left|\dth{S, prel} - 1.7\dth{P, prel}\right| > \SI{0.1}s
\end{equation}

correspond to unrealistic {\vpvsratio}s and are removed. The removed data points resulted from wrongly associated picks in the used catalog. \dth{P, prel}, \dth{S, prel} are recalculated by removing the median. In a next step, data points with

\begin{equation}
\sqrt{{\dth{P, prel}}^2 + \parent{\dth{S, prel}/R}^2} > \SI{0.35}{s}
\end{equation}

are additionally removed. This is in accordance with \cite{Dahm2013} and removes event pairs whose locations are far away from each other. Here, data points outside a circle (or ellipse for $R{\neq}1$) are removed to not affect the subsequent regression. Finally, \dth P, \dth S are calculated by removing the mean for each event pair (equation~\ref{eq:vpvs2}) and a robust orthogonal L1 regression is performed by testing different ratios with brute force (equation~\ref{eq:vpvs1}). For $R{=}1$ I obtain {\vpvsratio}s of 1.73, 1.70, 1.72, 1.71, 1.71 for the earthquake clusters \emph a, \emph b, \emph c, \emph d and \emph e.

\cite{Lin2007} argue that errors in the differential S wave travel times due to different take-off angles are theoretically larger by a factor $R{=}v\ind P/v\ind S$ than the corresponding errors for differential P wave travel times. They therefore suggest to determine the \vpvsratio iteratively and scale the S~wave differential travel times appropriately. With this approach using the same equations as above with $R{=}v\ind P/v\ind S$, velocity ratios of 1.68, 1.66, 1.69, 1.68, 1.69 are obtained for the clusters \emph a, \emph b, \emph c, \emph d and \emph e (see figure~\ref{fig:vpvs}). These estimates are slightly lower than the estimates for $R{=}1$, but in any case the \vpvsratio is similar for all analyzed earthquake clusters and consistent with previous observations \citep[e.g.][]{Malek2005}.

\catcode`\^^M=5



\end{document}